\documentclass[twocolumn,aps,showpacs,prb,tightenlines,amsmath,amssymb,superscriptaddress]{revtex4-1}
\usepackage{graphicx}
\usepackage{amssymb}
\usepackage{amsmath}
\usepackage{colordvi}
\usepackage{hyperref}

\newcommand{\bgreek}[1]{\mbox{\boldmath$#1$\unboldmath}}
\newcommand{\be}{\begin{equation}}
\newcommand{\ee}{\end{equation}}

\newcommand{\bea}{\begin{eqnarray}}
\newcommand{\eea}{\end{eqnarray}}

\newcommand{\ii}{{\rm i}}
\newcommand{\vep}{\varepsilon}
\newcommand{\ome}{\omega}

\newcommand{\up}{\uparrow}
\newcommand{\down}{\downarrow}
\newcommand{\ave}[1]{\langle #1\rangle}

\begin{document}

\title{Spin Susceptibility and Helical Magnetic Orders at the
  Edges/Surfaces of Topological Insulators Due to Fermi Surface Nesting}

\author{Jian-Hua Jiang}
\affiliation{Department of Condensed Matter Physics, Weizmann Institute of
  Science, Rehovot, 76100, Israel}
\author{Si Wu}
\affiliation{Department of Physics, University of Toronto, 60 St. George St., 
Toronto, ON M5S 1A7, Canada}

\date{\today}

\begin{abstract}
We study spin susceptibility and magnetic order at the edges/surfaces
of two-dimensional and three-dimensional topological insulators when
the Fermi surface is nested. We find that due to spin-momentum locking
as well as time-reversal symmetry, spin susceptibility at the nesting
wavevector has a strong {\em helical} feature. It follows then, a {\em
  helical} spin density wave (SDW) state emerges at low temperature
due to Fermi surface nesting. The helical feature of spin
susceptibility also has profound impact on the magnetic order in
magnetically doped surface of three dimensional topological
insulators. In such system, from the mean field Zener theory, we
predict a {\em helical} magnetic order.
\end{abstract}


\maketitle

\section{Introduction}

In the past few years, a new family of materials called topological
insulators (TIs) have been theoretically predicted
\cite{KaneMele2,BHZ,FuKaneMele,FuKane,MooreBalents,Roy}
and then experimentally observed.\cite{Konig,Hsieh,Xia,YLChen}
A TI has a energy gap in the bulk and gapless excitation 
in the edge/surface, which is due to the nontrivial band topology
and protected by time reversal symmetry. TIs possess a nontrivial
topological order, which distinguishes them from simple band
insulators. In a two dimensional (2D) TI, which is also known as a
quantum spin Hall insulator, the edge states form a helical
Luttinger liquid.\cite{WuBernevigZhang} 
The surface states of three dimensional (3D) TIs
form a ``helical metal'' with Dirac cone like spectrum
\cite{FuKaneMele,FuKane,Zhang,Hsieh,Xia,Hsieh2,FFT-STM}
(Specifically, in this paper, we are interested in a class of 3D TIs
where the surface state consists of a single Dirac
cone. Examples are Bi$_2$Te$_3$,\cite{Zhang,YLChen}
Bi$_2$Se$_3$,\cite{Zhang,Xia,Kuroda} Sb$_2$Te$_3$,\cite{Zhang,Hsieh3}
TlBiSe$_2$\cite{Kuroda2,TlBiSe}, and many other
materials\cite{TlBiTe,SYXu}. Due to large bandgap and high purity, these
materials have great potential for application and
scientific research.\cite{HasanKane}). An important feature is that
spin and momentum are closely correlated in the edge/surface states of
TIs, which leads to many unusual
effects\cite{Dai,GV,HasanKane,Qireview} and potential applications in
spintronics and quantum computation.\cite{spintronics}

Recently, it was found in Bi$_2$Te$_3$ that the as Fermi energy
increases from the Dirac point, the shape of Fermi surface gradually
changes from a circle, first to a hexagonal shape, and then to a
snowflake-like.\cite{YLChen,Alpich} This phenomenon was also found in
other 3D TIs with similar
structures.\cite{Kuroda,Hsieh3,TlBiSe,TlBiTe,SYXu}
This kind of band structure is theoretically reproduced by Fu from the
$k\cdot p$ theory.\cite{Fu} For a certain range of energies, the Fermi
surface is almost a hexagon, which leads to strong nesting at
three wavevectors and possible instability to the formation of SDW
states.\cite{Fu}

In this paper, we study spin susceptibility and magnetic order at the
edges/surfaces of TIs when the Fermi surface is nested. We find that
due to the one-to-one correspondence between spin state and
momentum (``spin-momentum locking'') as well as time reversal
symmetry, the spin susceptibility function at nesting wavevector has a
strong helical feature. It follows then, a {\em helical} SDW 
state emerges at low temperature due to Fermi surface nesting. We
present a mean field theory of the helical SDW state. The helical
feature of the spin susceptibility function also has profound impact
on the magnetic order in the magnetically doped surfaces of 3D TIs. In
such system, from the mean field Zener theory, we predict a
{\em helical} magnetic order.

\section{Hamiltonian, spectrum and eigenstates}

We consider the situation where Fermi surface exhibits strong nesting
feature. Examples are, the edge states in a 2D TI, and the surface state
in the 3D TI Bi$_2$Te$_3$ with Fermi energy in the range of [0.13,
  0.23]~eV where the Fermi surface is almost a hexagon
\cite{YLChen,Fu} [see Fig.~1]. The hexagonal shape of Fermi surface
also exists in the surface states of many other 3D TIs, such as
Bi$_2$Se$_3$,\cite{Liu,Kuroda} Sb$_2$Te$_3$,\cite{Fu,Liu}
TlBiSe$_2$\cite{Kuroda2,TlBiSe}, TlBiTe$_2$\cite{TlBiTe} and a
recently discovered large class of 3D TIs in
Ref.~\onlinecite{SYXu}. In the following we would call these materials the
``Bi$_2$Se$_3$ class''.

The Hamiltonian of the edge states in a 2D TI can be written as\cite{WuBernevigZhang}
\be
H_0 = \sum_{k\alpha\beta} v_0 k~ c^\dagger_{k\alpha} \sigma^z_{\alpha\beta}
c_{k\beta},
\label{1dh}
\ee
where $v_0$ is the Fermi velocity. The spin orientations of the
eigenstates, through proper choosing of the spin coordinates, have
been taken as up and down. The eigenenergies and eigenstates are
\be
\vep_{\pm}(k) = \pm v_0 |k|, \quad 
u_{\pm}(k) = \left(\begin{array}{c} \Theta(\pm k) \\ \Theta(\mp k) \end{array}\right),
\ee
where $\Theta$ is the Heaviside function. The nesting vector
is $Q=2k_F$.

Throughout this paper, we focus on the situation where the Fermi
energy $E_F$ is high and the temperature is low, so that states far
below the Fermi surface (such as those below the Dirac point) is
irrelevant. This regime is easy to achieve in the Bi$_2$Se$_3$ class
of 3D TIs, thanks to the large bandgap and Dirac velocity.

\begin{figure}[t]
\includegraphics[height=3.8cm]{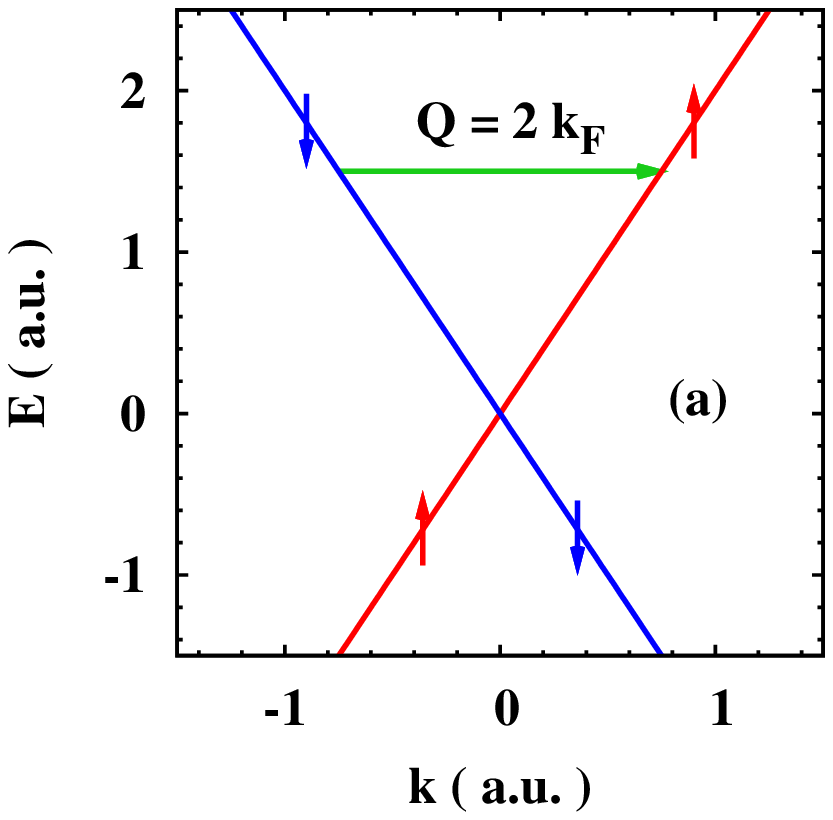}\includegraphics[height=3.8cm]{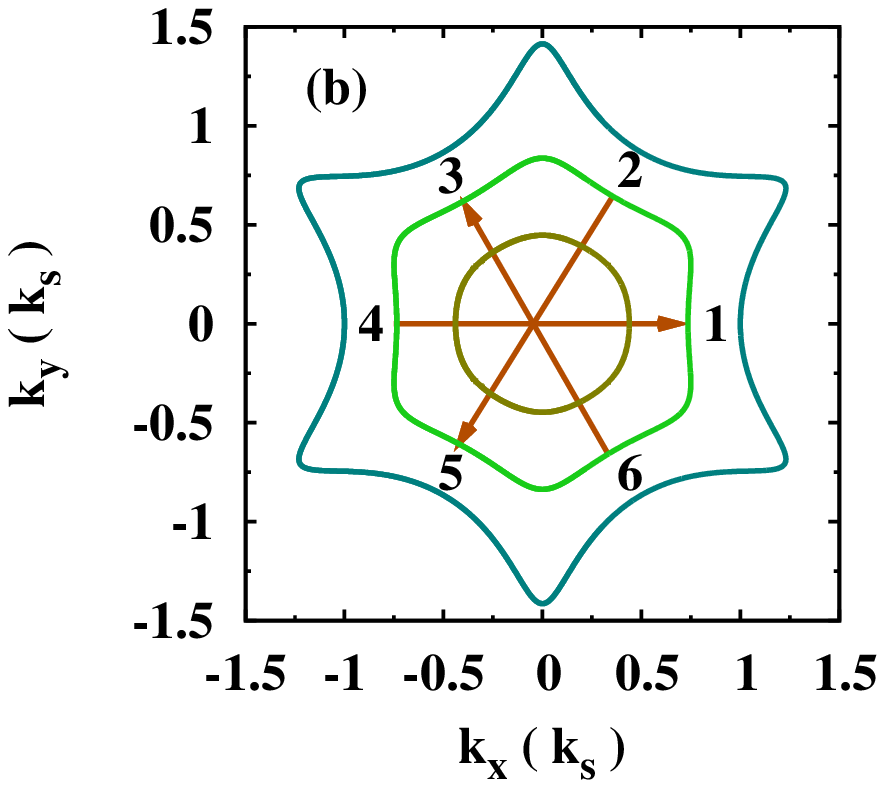}
\caption{(Color online) (a) Spectrum of the edge states of 2D TIs as well as
  schematic of the nesting wave-vector $Q=2k_F$. Nesting wavevector
  connects spin-down and spin-up states. (b) Energy contour of the
  surface states of 3D TIs for three typical energy, $0.2, 0.7,
  2$~$E_s$, as well as schematic of nesting wave-vectors for hexagonal
  Fermi surface. The numbers label the Fermi Arcs. There are three nesting
  wavevectors, ${\bf Q}_1$ from Arc 4 to Arc 1, ${\bf Q}_2$ from Arc 6
  to Arc 3, ${\bf Q}_3$ from Arc 2 to Arc 5.}
\end{figure}

The Hamiltonian for the surface states in 3D TI with a single Dirac
cone, is given in Refs.~\onlinecite{Fu,Liu}. Keeping only the dominant
terms, the Hamiltonian is (with $x$ and $y$ axes along
the $\Gamma-K$ and $\Gamma-M$ directions respectively),\cite{Fu,Liu}
\be
H_0 = v (k_x\sigma_y - k_y\sigma_x) +
\frac{\gamma}{2}(k_{+}^3+k_{-}^3)\sigma_z,
\label{h2d}
\ee
where $k_{\pm}=k_x\pm \ii k_y$. The eigenenergy and spin states are\cite{Liu}
\bea
\vep_{\pm}({\bf k}) &=& \pm \sqrt{v^2k^2+\gamma^2 k^6
  \cos^23\theta_k}, \label{disper} \\
u_{\pm}({\bf k}) &=& \frac{1}{\sqrt{A({\bf k})}}\left(\begin{array}{c}
  vk(\sin\theta_k+\ii\cos\theta_k) \\  \vep_{\pm}({\bf k}) - \gamma
  k^3\cos3\theta_k \end{array}\right),
\eea
with ${\bf k}=k(\cos\theta_k,\sin\theta_k)$, $A({\bf k})=v^2k^2 + (\vep_+({\bf k}) - \gamma
k^3\cos3\theta_k)^2$. The Hamiltonian is time-reversal invariant.
The last term reduces the symmetry from $C_{\infty v}$ to $C_{3v}$.
The Hamiltonian describes the surface state of the Bi$_2$Se$_3$ class
of 3D TIs.\cite{Fu,Liu} It is instructive to rescale the Hamiltonian
with,\cite{Fu} 
\be
H_0 = E_s H_0^{\prime},\quad k = k_sk^{\prime},
\ee
where
\be
E_s=v\sqrt{v/\gamma}, \quad k_s=\sqrt{v/\gamma}.
\ee
Then the Hamiltonian becomes
\be
H_0^\prime = (k_x^\prime\sigma_y - k_y^\prime\sigma_x) +
\frac{1}{2}(k_{+}^{\prime 3}+k_{-}^{\prime 3})\sigma_z.
\ee
Obviously, the physical properties are universal in these materials,
with exact quantitative scaling by properly recovering the
dimension via $E_s$ and $k_s$ within the above model. The parameters,
$v$, $\gamma$, $E_s$ and $k_s$ of several 3D TIs inferred from
experiments are listed in Table~I.

\begin{table}[htbp]
  \caption{Hamiltonian parameters of the surface states in 3D TIs with
    hexagonal symmetry. ($^a$ from Ref.~\onlinecite{Fu}, $^b$ from
    Ref.~\onlinecite{Kuroda}, $^c$ inferred from 
    Ref.~\onlinecite{TlBiSe}, $^d$ inferred from Ref.~\onlinecite{SYXu}.)} 
  \begin{center}
    \begin{tabular}{lllllllll} \hline \hline
      &\mbox{} & $v$ (eV\AA) &\mbox{} & $\gamma$ (eV\AA$^{3}$) &\mbox{}
      & $E_s$ (eV) &\mbox{} & $k_s$ (\AA$^{-1}$) \\ \hline
      Bi$_2$Te$_3$$^a$: &\mbox{} & 2.55 &\mbox{} & 250 &\mbox{} & 0.26
      &\mbox{} & 0.1 \\
      Bi$_2$Se$_3$$^b$: &\mbox{} & 3.55 &\mbox{} & 128 &\mbox{} & 0.59
      &\mbox{} & 0.17 \\
      TlBiSe$_2$$^c$: &\mbox{} & 3.1 &\mbox{} & 182 &\mbox{} & 0.4
      &\mbox{} & 0.13 \\
      GeBi$_2$Te$_4$$^d$: &\mbox{} & 2.37 &\mbox{} & 99 &\mbox{} & 0.37
      &\mbox{} & 0.15 \\
      Bi$_2$Te$_2$Se$^d$: &\mbox{} & 7.25 &\mbox{} & 580 &\mbox{} & 0.81
      &\mbox{} & 0.11 \\
  \hline \hline
\end{tabular}
\end{center}
\end{table}

If the Fermi energy is in the range of $0.5E_s\le E_F\le 0.9 E_s$, the Fermi
surface is almost a hexagon, which exhibits strong nesting feature.\cite{Fu}
There are three nesting wave-vectors
\bea
{\bf Q}_1 &=& 2k_0(1,0),\quad\quad{\bf Q}_2 =
2k_0(-\frac{1}{2},\frac{\sqrt{3}}{2}),\nonumber \\ \quad\quad {\bf Q}_3 &=&2k_0
(-\frac{1}{2},-\frac{\sqrt{3}}{2}),
\label{Q123}
\eea
where $k_0$ is determined by $\sqrt{v^2k_0^2+\gamma^2 k_0^6}=E_F$.

At higher Fermi energy, the Fermi surface is distorted to
snowflake-like and new nesting wave-vectors
emerges.\cite{Fu,Alpich} However, at such high Fermi energy
usually the bulk conduction band is also occupied, which complicates
the situation and is not interested in this paper.

The electron-electron interaction consists of the long-range Coulomb
interaction and the short-range Hubbard interaction. The former does
not affect the spin susceptibility [to the random phase approximation
(RPA)\cite{note0}] and is hence ignored. The onsite Hubbard
interaction is written as
\bea
H_U &=& - \frac{U}{3} \sum_{{\bf q}} \bgreek{\sigma}({\bf q})\cdot \bgreek{\sigma}({\bf -q}),
\eea
where $\bgreek{\sigma}({\bf q}) = \sum_{{\bf
    k}\alpha\beta}c^\dagger_{{\bf
    k}\alpha}\bgreek{\sigma}_{\alpha\beta}c_{{\bf k}+{\bf q}\beta}$
with $\bgreek{\sigma}$ being Pauli matrices vector and $\alpha,\beta$
being spin indices. For positive Hubbard $U$ (repulsive interaction),
Fermi surface nesting leads to the SDW instability and transition into
the SDW state at sufficient low temperature. Whereas, for negative $U$
(attractive interaction), it leads to the charge density wave (CDW)
instability and the emergence of CDW state. Here we assume, as in most
cases, $U>0$.

\section{Spin susceptibility and SDW instability}

\subsection{Edge states of 2D TIs}

The linear spin susceptibility function is
\be
\chi_{\mu\nu}(q, \ome) = \ii \int_0^{\infty}\hspace{-0.2cm}dt \ave{[\sigma_{\mu}(
      q,t), \sigma_{\nu}(-q,0)]}e^{\ii t(\ome+\ii 0^{+})},
\ee
where $\mu,\nu=(x,y,z)$ or $(\pm,z)$ with
$\sigma_{\pm}=\frac{1}{2}(\sigma_x\pm\ii\sigma_y)$. In the edge of a
2D TI, at the nesting vector $Q=2k_F$, the only nonzero terms are
$\chi_{+-}(-Q,\ome)$ and $\chi_{-+}(Q,\ome)$ as the nesting wavevector
connects spin-down and spin-up states. This helical feature of the
spin susceptibility function is due to spin-momentum locking
as well as time reversal symmetry. The two susceptibility 
function are actually related by
$[\chi_{+-}(-Q,-\ome)]^\ast=\chi_{-+}(Q,\ome)$. In the absence of
interaction, the spin susceptibility is
\be
\chi^0_{-+}(Q,\ome) =
 \sum_k \frac{ n_F(\xi_{k+Q\up})-n_F(\xi_{k\down}) }{\ome -\xi_{k+Q\up}
  +\xi_{k\down}+i0^+},
\label{chidiver}
\ee
where $\xi_{k\down}=\vep_{k\down}-E_F$ and $n_F$ is the Fermi
distribution. Including the Hubbard interaction within RPA, one gets
\be
\chi_{-+}(Q,\ome) = \frac{\chi^0_{-+}(Q,\ome)}{1-U\chi^0_{-+}(Q,\ome)}.
\label{chi1}
\ee
Due to Fermi surface nesting, the spin susceptibility
$\chi^0_{-+}(Q,0)$ is divergent at low temperature.
Actually, directly from Eq.~(\ref{chidiver}), one can show that
$\chi^0_{-+}(Q,0) \approx \frac{1}{\pi v} \log(E_F/k_BT)$. That is, the
spin susceptibility function is logarithmically divergent with
decreasing temperature. This signals the SDW instability. The feature
that only $\chi^0_{-+}(Q,0)$ diverges indicates a {\em helical} SDW
order.

The above treatment based on the Fermi-liquid theory is of course
invalid for one-dimensional electron system, but it sheds some light
on the problem. In the following, we analyze the problem via the
bosonization theory.

Following Wu et al., \cite{WuBernevigZhang} the bosonized Hamiltonian 
in the presence of Umklapp scattering can be expressed as
\bea
H&=&\frac{1}{2\pi}\int dx\left[uK\left(\nabla\theta(x)\right)^2+\frac{u}{K}\left(\nabla\phi(x)\right)^2\right]
\nonumber \\ &&\mbox{}~~ +\frac{g_u}{2(\pi a)^2}\cos(4\phi(x)),
\eea
where the bosonized fermion fields are
\be
\psi_{R\up}(x)=\frac{e^{ik_Fx}}{\sqrt{2\pi
    a}}e^{-i\phi_R(x)},\quad
\psi_{L\down}(x)=\frac{e^{-ik_Fx}}{\sqrt{2\pi a}}e^{i\phi_L(x)},
\ee
with $\phi=(\phi_R+\phi_L)/2$, $\theta=(\phi_R-\phi_L)/2$. $K$ the
Luttinger parameter. $u$ is the renormalized Fermi velocity. $g_u$
is the Umklapp scattering strength. $a$ is a short-distance
cutoff. Due to symmetry reasons, the only possible instabilities are
SDW and singlet superconductivity (SC). This 
is because CDW and triplet SC instabilities pair particles with the
same spin, i.e., terms like $\psi_{R\up}^{\dagger}\psi_{L\up}$
for CDW and $\psi_{R\up}^{\dagger}\psi_{L\up}^{\dagger}$ for
triplet SC, are impossible. The bosonized form of spin operators are
\bea
\sigma_+(x)=\psi_{R\up}^{\dagger}(x)\psi_{L\down}(x)=\frac{e^{-i2k_Fx}}{2\pi a}
e^{2i\phi(x)},\nonumber\\
\sigma_-(x)=\psi_{L\down}^{\dagger}(x)\psi_{R\up}(x)=\frac{e^{i2k_Fx}}{2\pi a}
e^{-2i\phi(x)}.
\eea
From standard bosonization theory, \cite{Giamarchi} the Umklapp term becomes
relevant when $K<1/2$. Then RG flow will go to a
strong coupling fixed point, $g_u\rightarrow \infty$, and the $\phi$ field will become 
ordered. Depending on the sign of $g_u$, the ordered value of $\phi$ is
\bea
\langle\phi\rangle&=&\frac{\pi}{4}+\frac{2n\pi}{4},\ \ \ \ \ g_u>0,\nonumber	\\
\langle\phi\rangle&=&0+\frac{2n\pi}{4},\ \ \ \ \ \ g_u<0.	
\eea
This signifies a true phase transition. Then the spin operators, which have zero expectation 
values in the non ordered phase, also acquire nonzero average value across the transition,
\bea
\langle\sigma_x\rangle&=&\langle \sigma_++\sigma_-\rangle=\frac{2}{(2\pi a)^2}
\cos(2k_Fx-\langle\phi\rangle),\nonumber	\\
\langle\sigma_y\rangle&=&\frac{1}{i}\langle \sigma_+-\sigma_-\rangle=\frac{-2}{(2\pi a)^2}
\sin(2k_Fx-\langle\phi\rangle),
\eea
which shows helical structure and is consistent with the mean field
result. We note that very recently a similar calculation has been
carried out by Kharitonov,\cite{Kharitonov} considering helical
Luttinger liquid in the proximity to a ferromagnet, which also agrees
with our result.

\subsection{Surface states of 3D TIs}

\paragraph{General considerations} On the surface of a 3D TI, the nesting vector will connect states
which are not Kramers pairs and hence their spin states are {\em not}
antiparallel. As a consequence, the spin susceptibility is finite in
all directions. The free spin susceptibility function in this case is
\be
\chi^0_{\mu\nu}({\bf Q},0) = \sum_{\bf k} \frac{
  n_F(\xi_{{\bf k+Q}+})-n_F(\xi_{{\bf k}+}) }{\xi_{{\bf k}+}-\xi_{{\bf
      k+Q}+}+i 0^+} g_{\mu}({\bf k},{\bf Q})g^{\ast}_{\nu}({\bf
  k},{\bf Q})
\label{chi-3d}
\ee
where 
\be
g_{\mu}({\bf k},{\bf Q})=\langle u_{+}({\bf k})|\sigma_{\mu}|u_{+}({\bf
  k+Q})\rangle
\ee
with $\mu,\nu=(x,y,z)$. One can note that,
$g_{\mu}({\bf k},{\bf Q})g^{\ast}_{\nu}({\bf k},{\bf Q})$ is a {\em
  bilinear} tensor, where $g_{\mu}({\bf k},{\bf Q})$ is its
eigenvector with eigen-value 
\be
\kappa_g({\bf k},{\bf Q})=\sum_{\mu}|g_{\mu}({\bf
  k},{\bf Q})|^2.
\ee
Moreover, if
\be
g_0({\bf k},{\bf Q}) = \langle u_{+}({\bf k})|u_{+}({\bf k+Q})\rangle,
\ee
and 
\be
\chi^0_{c}({\bf Q},0) = \sum_{\bf k} \frac{
  n_F(\xi_{{\bf k+Q}+})-n_F(\xi_{{\bf k}+}) }{\xi_{{\bf k}+}-\xi_{{\bf
      k+Q}+}+i 0^+} |g_0({\bf k},{\bf Q})|^2,
\ee
is the charge susceptibility, then
\be
|g_0|^2 + \kappa_g({\bf k},{\bf Q}) \equiv 2, \quad 1\le \kappa_g({\bf k},{\bf Q}) \le 2.
\label{kg}
\ee
and a ``{\em complementary relation}'' between the spin- and charge-
susceptibility,
\be
\sum_\mu\chi^0_{\mu\mu}({\bf Q},0) + \chi^0_c({\bf
  Q},0)  = \chi^0({\bf Q},0)
\ee
with $\chi^0({\bf Q},0) \equiv \sum_{\bf k} \frac{
  n_F(\xi_{{\bf k+Q}+})-n_F(\xi_{{\bf k}+}) }{\xi_{{\bf k}+}-\xi_{{\bf
      k+Q}+}+i 0^+}$, due to spin-momentum locking. We then introduce
the spin density operator
\be
\sigma_g({\bf k},{\bf Q}) =
\frac{1}{\kappa_g({\bf k},{\bf Q})} \sum_{\mu,\alpha\beta}
g_\mu^{\ast}({\bf k},{\bf Q})c^\dagger_{{\bf k}\alpha}
\sigma_\mu^{\alpha\beta}c_{{\bf k}+{\bf Q}\beta}.
\ee
One can show that
\be
\langle u_{+}({\bf k})|\sigma_{g}({\bf k},{\bf
  Q})|u_{+}({\bf k+Q})\rangle = 1.
\ee
And, any spin density operator
\be
\sigma_f({\bf k},{\bf Q}) = \sum_{\mu,\alpha\beta} f_\mu^\ast({\bf
  k}) c^\dagger_{{\bf k}\alpha} \sigma_\mu c_{{\bf k}+{\bf
    Q}\beta},
\ee
which is perpendicular to $\sigma_g$, i.e., $\sum_\mu g_\mu^\ast({\bf
  k}) f_\mu({\bf k}) = 0$, has 
\be
\langle u_{+}({\bf
  k})|\sigma_{f}({\bf k},{\bf Q})|u_{+}({\bf k+Q})\rangle =
0.
\ee
Therefore, only one spin density susceptibility is nonzero (if
only ${\bf k}$ and ${\bf k}+{\bf Q}$ states are concerned), which is
defined by $\sigma_g({\bf k},{\bf Q})$. This intriguing feature is
due to spin-momentum locking.

\paragraph{Helical feature}
Consider, if ${\bf k}$ and ${\bf k}+{\bf Q}$ are time reversal pair
states, of which spin orientations are opposite, the spin density
operator $\sigma_g({\bf k},{\bf Q})$ is helical, as we learn from
the case in the edges of 2D TIs. Unfortunately, here $g_\mu({\bf
  k},{\bf Q})$ is ${\bf k}$ dependent. Besides, only for some 
very special ${\bf k}$ states, such as ${\bf k}_0=(-k_0,0)$ and ${\bf
  k}_0+{\bf Q}_1=(k_0,0)$, the two states are time reversal pairs. If
the contributions of all ${\bf k}$ states are summed, the helical
feature may be smeared out.

To study this case, we calculate the free spin susceptibility
function $\chi_{\mu\nu}^0({\bf Q_1},0)$ numerically for $E_F\in[0.5,
  0.9]$~$E_s$ where the Fermi surface is hexagonal (${\bf
  Q}_1=(2k_0,0)$ [see Eq.~(\ref{Q123})] connects Fermi Arc 
4 and Arc 1 [see Fig.~1(b)]). Our results are presented in
Fig.~2. From the calculation, we find that, though the exact helical
spin susceptibility is ruled out, the spin susceptibility function
still has {\em strong helical} feature. Specifically, one of the eigenvalues
of the spin susceptibility tensor $\chi_{\mu\nu}^0({\bf Q_1},0)$ is much
larger than the other two [Fig.~2(b)], which corresponds to a spin
density response with helicity {\em very close to unity}
[Fig.~2(c)]. The helical spin rotating axis is indicated in Fig.~2(a)
as ${\bf n}_g$, which lies in the $y$-$z$ plane with an angle $\theta_n$
[Fig.~2(c)] between $y$-axis.

\begin{figure}[bth]
 \begin{minipage}[h]{0.45\linewidth}
   \vskip -0.1cm
   \centerline{\includegraphics[height=3.1cm]{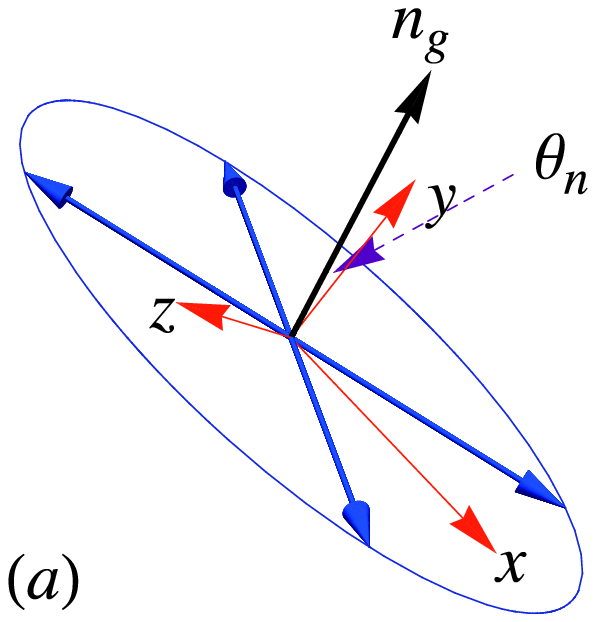}}
  \end{minipage}\hfill
  \begin{minipage}[h]{0.55\linewidth}
    \centerline{\includegraphics[height=2.7cm]{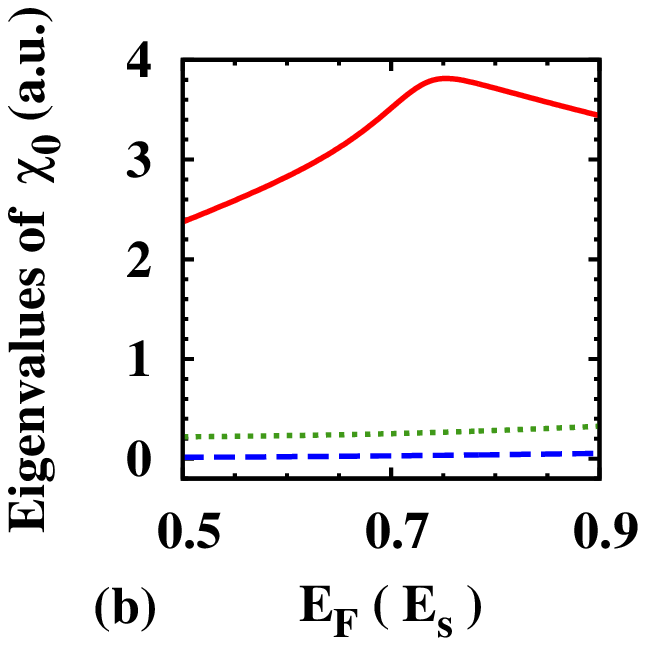}}
    \centerline{\includegraphics[height=2.8cm]{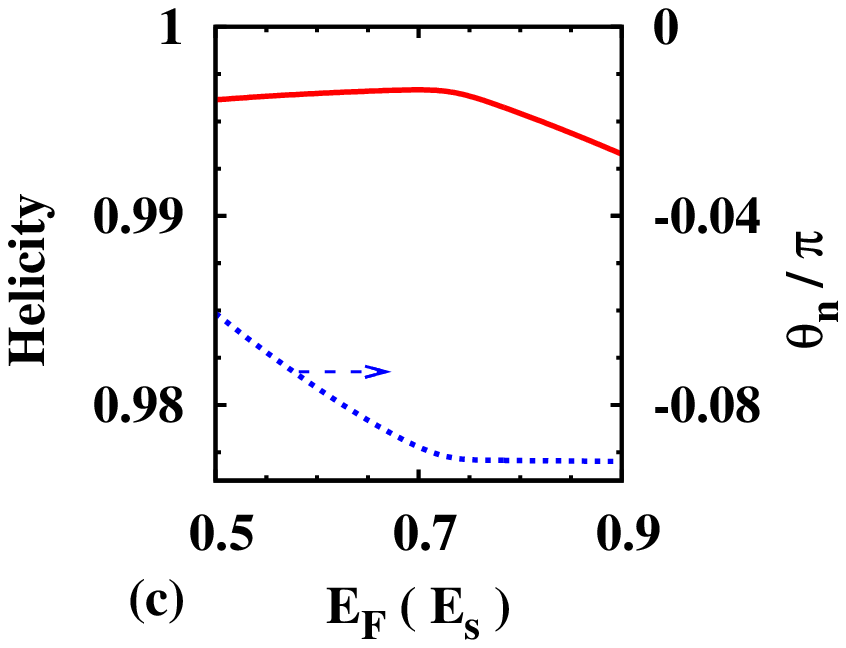}}
  \end{minipage}\hfill
\caption{(Color online) (a) Schematic of the spin density response
  corresponding to the largest eigen-value of the spin susceptibility
  tensor $\hat{\chi}^0({\bf Q}_1,0)$. ${\bf n}_g$ is the helical
  rotating axis. The blue arrows indicate spins rotating around the
  ${\bf n}_g$ axis. ${\bf n}_g$ lies in the $y$-$z$ plane with an
  angle $\theta_n$ between $y$ axis. (b) Eigenvalues (positive
  definite) of the tensor $\hat{\chi}^0({\bf Q}_1,0)$ as function of
  Fermi energy (in unit of $E_s$). (c) The helicity and the angle $\theta_n$ as
  function of Fermi energy. All data are calculated at $k_BT=1.3\times
  10^{-3} E_s$.}
\end{figure}

\paragraph{Physical explanations}
To understand the above results, let's consider the problem that
how the spin susceptibility function gets maximized at ${\bf Q}_1$.
The spin susceptibility function for an arbitrary spin density operator,
$\sigma_f({\bf Q}_1)=\sum_{\mu}f_{\mu}^{\ast}\sigma_\mu({\bf
  Q}_1)$ [$\sum_{\mu}|f_\mu|^2=1$ (normalized), and $\sigma_\mu({\bf
  Q}_1)=\sum_{{\bf k},\alpha\beta}c^\dagger_{{\bf
    k}\alpha}\sigma_{\mu}^{\alpha\beta}c_{{\bf k}+{\bf Q}_1\beta}$], is
\bea
&&\hspace{-0.7cm}\chi^0_f({\bf Q}_1,0) = \sum_{\bf k} \frac{ n_F(\xi_{{\bf k}+}) -
  n_F(\xi_{{\bf k+Q}_1+}) }{\xi_{{\bf k+Q}_1+}-\xi_{{\bf k}+}}
\nonumber
\\ &&\hspace{-0.1cm} \times \Big|\sum_{\mu}f_{\mu}^{\ast} \frac{g_{\mu} ({\bf k},{\bf
     Q}_1)}{\sqrt{\kappa_g({\bf k},{\bf Q}_1)}}\Big|^2 \kappa_g({\bf k},{\bf Q}_1) ,
\label{opchi}
\eea
The spin susceptibility gets contribution mainly from the states
satisfying the nesting condition and from the vicinity. The above
equation also indicates that the contribution is proportional to
the factor $\kappa_g({\bf k},{\bf Q}_1)$. It is easy to show that
for the special wavevector ${\bf k}_0=(-k_0,0)$, the nesting condition
is exactly satisfied, $\xi_{{\bf k}_0+}=\xi_{{\bf k}_0+{\bf
    Q}_1+}=0$. Moreover, from (denote $g^0_{\mu}\equiv
g_{\mu}({\bf k}_0,{\bf Q}_1)$),
\be
g^0_x=\ii,\quad g^0_y=\frac{-\gamma
  k_0^3}{\sqrt{v^2k_0^2+\gamma^2k_0^6}},\quad
g^0_z=\frac{-vk_0}{\sqrt{v^2k_0^2+\gamma^2k_0^6}} ,
\label{g0}
\ee
one finds that the magnitude $\kappa_g({\bf k},{\bf Q}_1)$ is largest
at ${\bf k}_0=(-k_0,0)$ with $\kappa_g({\bf k}_0,{\bf Q}_1)=2$. As
${\bf k}_0$ is at the center of the Fermi Arc 4, where the main
contribution to spin susceptibility comes, the overlapping factor
$\Big|\sum_{\mu}f_{\mu}^{\ast} \frac{g_{\mu} ({\bf
    k},{\bf Q}_1)}{\sqrt{\kappa_g({\bf k},{\bf Q}_1)}}\Big|^2$ is also
very large, if $f_{\mu}^{\ast}=g^{0
  \ast}_\mu/\sqrt{\kappa_g({\bf k}_0,{\bf Q}_1)}$. Calculation indicates that the
overlapping factor is very close to unity in the vicinity of Fermi Arc
4. Therefore, the spin density operator $\sigma_f$ which maximizes the
spin susceptibility function should be $\sigma_g^0({\bf
  Q}_1)=\sum_{\mu} g^{0 \ast}_\mu \sigma_{\mu}({\bf
  Q}_1)/\sqrt{\kappa_g({\bf k}_0,{\bf Q}_1)}$. This observation is
very close to the truth, except that it ignores some delicate
part. Although the factor $\frac{ n_F(\xi_{{\bf k}+}) -
  n_F(\xi_{{\bf k+Q}_1+}) }{\xi_{{\bf k+Q}_1+}-\xi_{{\bf k}+}}$ is
peaked at ${\bf k}_0=(-k_0,0)$ (as $\xi_{{\bf k}_0+}=\xi_{{\bf
    k}_0+{\bf Q}_1+}=0$), the dispersion, however, is {\em asymmetric}
around ${\bf k}_0$ along the $k_x$ direction, as the spectrum is {\em
  nonlinear} [see Eq.~(\ref{disper}) and Fig.~1]. Therefore, the spin
density operator which maximize the spin susceptibility,
denoted as $\sigma_G$, deviates slightly from
$\sigma_g^0$. Nevertheless, $\sigma_G$ is very close to
$\sigma_g^0$ and keeps most of the features of $\sigma_g^0$,
especially the {\em helical feature}. This feature even persists to
{\em room temperature}, thanks to the large bandgap and Dirac velocity
in the Bi$_2$Se$_3$ class of 3D TIs.

One can write $\sigma_G({\bf Q}_1)=\sum_{\mu} G_{1\mu}^\ast
\sigma_\mu({\bf Q}_1)$, where
\be
G_{1\mu}\approx \frac{1}{\sqrt{2}}g^0_{\mu}~~,\quad\quad
\sum_\mu|G_{1\mu}|^2=1~.
\label{Gmu}
\ee
From Eq.~(\ref{g0}), we know that $G_{1x}$ is a pure imaginary number,
whereas $G_{1y}$ and $G_{1z}$ are real numbers. Also, $iG_{1x}, G_{1y},
G_{1z}<0$. Hence the helical spin rotation axis is 
\be
{\bf n}_g=(0,\cos\theta_n,\sin\theta_n)
\ee
with $\theta_n= {\rm Arctan}(G_{1z}/G_{1y})-\pi/2$, agreeing with the
results in Fig.~2. The helicity is $2\sqrt{G_{1y}^2+G_{1z}^2}|G_{1x}|$. From
Eqs.~(\ref{Gmu}) and (\ref{g0}), one can see that the helicity is
indeed close to unity.

\paragraph{Systematic results}
In Fig.~3(a), we plot the largest eigenvalue of the spin
susceptibility tensor, $\chi_G^0({\bf Q},0)$, as function of
$|{\bf Q}|$ with ${\bf Q}$ along the $x$-direction. It is
seen that the spin susceptibility function has a strong peak at the
nesting wave-vector ${\bf Q}_1$, especially when $E_F=0.7 E_s$ where
the Fermi surface is almost a perfect hexagon. We also plot the spin
susceptibility function at a higher temperature $T=100$~K (blue dotted
curve) with chemical potential $0.7 E_s$. We find that the nesting and
the helical features persist to high temperature. And the largest
eigenvalue of the spin susceptibility tensor is still much larger than
the other two for ${\bf Q}\sim {\bf Q}_1$, at high temperature.

We present a two-dimensional plot of $\chi_G^0({\bf Q},0)$ in Fig.~3(b).
It is seen that the spin susceptibility function peaks at regions
close to the nesting wavevectors $\pm {\bf Q}_i$ $i=(1,2,3)$. Spin
susceptibility is small at both small and large ${\bf Q}$. This  is
quite different from the situation in a normal two-dimensional
electron system, where spin susceptibility at small ${\bf Q}$ is
large.

\begin{figure}[htb]
\includegraphics[height=3.4cm]{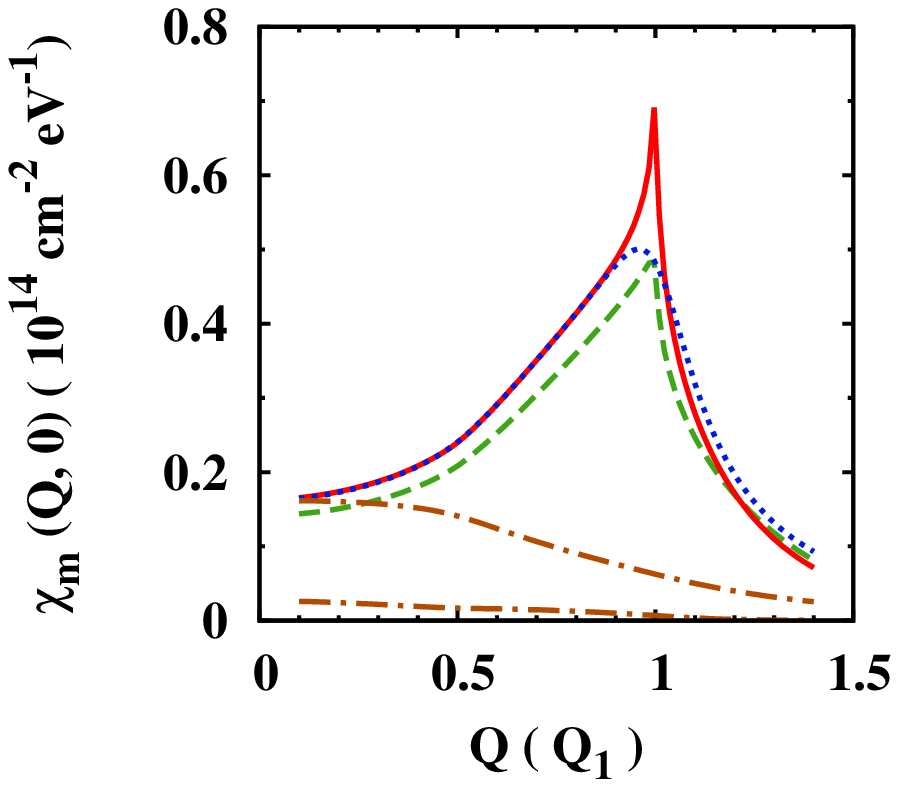}\includegraphics[height=3.4cm]{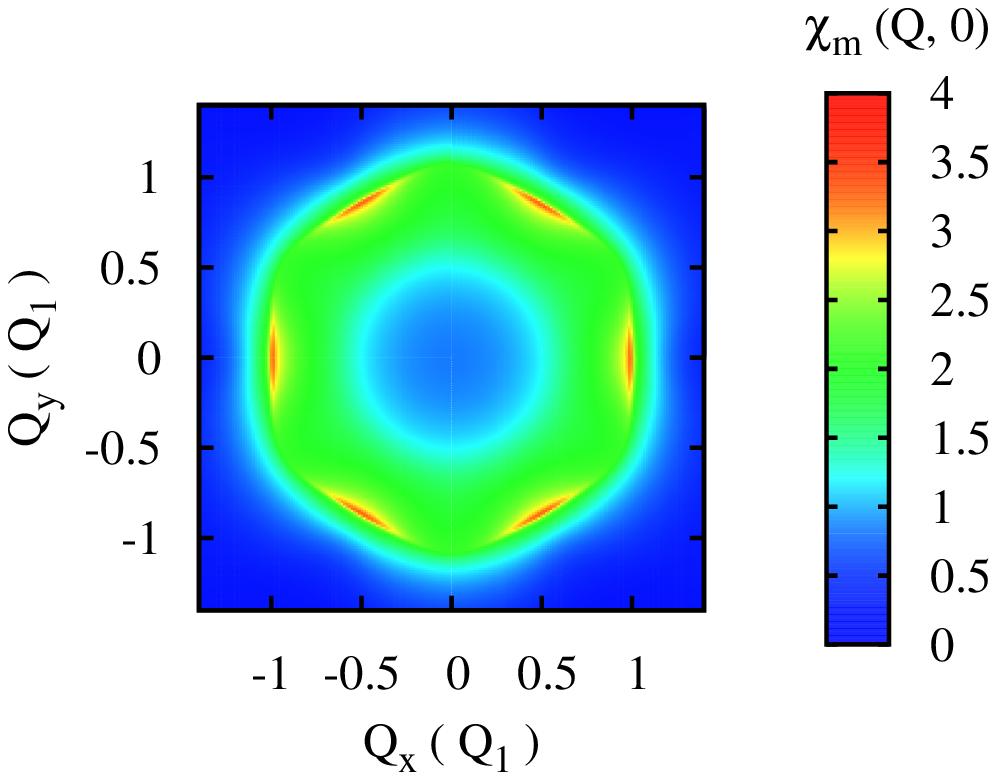}
\caption{(Color online) (a) The largest eigenvalue of spin
  susceptibility tensor, $\chi_G^0({\bf Q},0)$, as function of the
  wave-vector ${\bf Q}=(Q,0)$ along {\em $x$-direction}.
  Note that $Q$ is in unit of the nesting wavevector $Q_1=|{\bf
    Q}_1|=2k_0$. The red full (green dashed) curve correspond to the
  case with $E_F=0.7 E_s$ ($0.5 E_s$) and temperature $T=4$~K. The
  blue dotted curve corresponding to the case with chemical potential
  $0.7$~$E_s$ and $T=100$~K. The dot-dashed curves denote other two
  eigen-values of $\hat{\chi}^0$ for the $T=100$~K case. The
  parameters of Bi$_2$Te$_3$ [see Table~I] are used.\cite{Fu} (b)
  $\chi_G^0({\bf Q},0)$ as function of ${\bf Q}$ for $E_F=0.7 E_s$,
  $k_BT=1.3\times 10^{-3} E_s$. $\chi_G^0$ is in unit of $0.05(k_s^2/E_s)$.} 
\end{figure}

\paragraph{Interaction correction}

Then by including the Hubbard interaction within the RPA approximation,
one gets
\be
\chi_G({\bf Q}_1,0) = \frac{\chi_G^0({\bf Q}_1,0)}{1-a_GU\chi_G^0({\bf
    Q}_1,0)}.
\ee
Here 
\be
\chi_G({\bf Q}_1,0)=\ii
\int_0^{\infty}\hspace{-0.2cm}dt \ave{[\sigma_{G}({\bf Q}_1,t),
    (\sigma_{G}({\bf Q}_1,0))^\dagger]}e^{\ii t(0+\ii 0^{+})}.
\ee
\be
a_G = \frac{1}{2}\Big[1+\big|\sum_\mu G_{1\mu}^2\big|^2\Big],\quad
\frac{1}{2}\le a_G\le 1.
\label{a_G}
\ee
Here the awkward factor $a_G$ is due to the
factor that the spin density $\sigma_G({\bf Q}_1)$ is not properly
normalized. The normalized spin density is
\be
\tilde{\sigma}_G({\bf Q}_1)=\sqrt{a_G}\sum_{\mu}G_{1\mu}^\ast \sigma_\mu({\bf
  Q}_1) .
\label{sig-G}
\ee
It is more obvious to see this through two specific cases: i) If
$\{G_{1\mu}\}$ is a real vector, e.g., $(1,0,0)$, then $a_G=1$. This is
the non-helical case. The normalized spin density is just
$\tilde{\sigma}_G=\sigma_z$. ii) If $\{G_{1\mu}\}=\frac{1}{\sqrt{2}}(1,\ii,0)$,
then $a_G=\frac{1}{2}$, which is the helical case. The normalized spin
density is then
$\tilde{\sigma}_G=\sigma_{-}=\frac{1}{2}(\sigma_x-\ii\sigma_y)$. The 
expectation values of both $\sigma_z$ and $\sigma_{-}$ are less than
or equal to unity, which signals the normalization.

Accordingly, hereafter we use the normalized spin susceptibility 
function, 
\be
\tilde{\chi}_G \equiv a_G\chi_G.
\label{chi-G}
\ee
One then gets
\be
\tilde{\chi}_G({\bf Q}_1,0) = \frac{\tilde{\chi}_G^0({\bf Q}_1,0)}{1-U\tilde{\chi}_G^0({\bf
    Q}_1,0)},
\ee
which indicates the SDW instability at low temperature.

\paragraph{$\hat{\chi}$ at other nesting wave-vectors}
From the symmetry of the system, the spin susceptibility at other nesting
wave-vectors can be obtained. Due to the $C_3$ symmetry,
$\hat{\chi}({\bf Q_2},0) = {\cal P}^\dagger\hat{\chi}({\bf Q_1},0){\cal P}$, and
$\hat{\chi}({\bf Q_3},0) = {\cal P}^\dagger\hat{\chi}({\bf
  Q_2},0){\cal P}$ with ${\cal P}$ being the transformation matrix for
rotation around the $z$-axis by $2\pi/3$, i.e., the $C_3$ operator.
The spin density operators,
\be
\tilde{\sigma}_G({\bf Q}_i)=\sqrt{a_G}\sum_{\mu} G_{i\mu}^\ast
\sigma_\mu({\bf Q}_i),\label{Gi}
\ee transform as (${\Theta}$ denotes time-reversal
operation) 
\bea
&&C_3:\quad \tilde{\sigma}_G({\bf Q}_i)\to \tilde{\sigma}_G({\bf Q}_{i+1})\nonumber\\
&&{\Theta}:\quad \tilde{\sigma}_G({\bf Q}_i)\leftrightarrow -[\tilde{\sigma}_G({\bf
    Q}_i)]^\dagger.
\label{Gsym}
\eea
The complex vectors $\{G_{i\mu}\}$ transform as
\be
{\cal P}^\dagger\{G_{i\mu}\}=\{G_{i+1\mu}\}.
\ee

\section{Mean field theory of the helical SDW state}

\subsection{Edges of 2D TIs}
The mean spin density in the helical SDW state is
\be
\ave{\sigma_x}= 2S_0\cos(Qz),\quad \ave{\sigma_y} = -2S_0\sin(Qz),\nonumber
\ee
where $S_0>0$ (by properly choosing the coordinate) is the amplitude.
It is noted that only the helical SDW with {\em negative} helicity
(along $z$-axis) exists. This is due to the unique property of
spin-momentum locking in TIs. If $v_0$ in Eq.~(\ref{1dh}) is negative,
then only the helical SDW with {\em positive} helicity exists.

The properties of the ground state and quasi-particles in
the helical SDW state can be explored via the mean field
theory. Ignoring unimportant terms, the mean field Hubbard interaction
is
\bea
H_U &=& - 2U [\sigma_+(-Q)\ave{\sigma_{-}(Q)} +
  \ave{\sigma_{+}(-Q)}\sigma_-(Q)\nonumber\\
  && ~ -
  \ave{\sigma_{+}(-Q)}\ave{\sigma_{-}(Q)} ],
\eea
with $\ave{\sigma_{-}(Q)} = \ave{\sigma_{+}(-Q)} = S_0$. The mean field
Hamiltonian is then
\bea
&&\hspace{-0.54cm} H_{\rm MF} =\nonumber \\
&&\hspace{-0.32cm}\sum_p{}^{\prime}\left(\hspace{-0.1cm}\begin{array}{cc} c^\dagger_{p\down} &
  c^\dagger_{p+Q\up}  \end{array}\hspace{-0.1cm}\right) \left(\begin{array}{cc}
    \hspace{-0.1cm} \xi_{p\down} & \hspace{-0.1cm} B \\ \hspace{-0.1cm}
     B & \hspace{-0.1cm} \xi_{p+Q\up}
    \end{array}\hspace{-0.1cm}\right)\left(\hspace{-0.1cm} \begin{array}{c}
  c_{p\down} \\ c_{p+Q\up} \end{array}\hspace{-0.1cm}\right)
+ 2US_0^2,
\eea
where $B=-2US_0$. $p=k+k_F$ and
$\xi_{p\down}=-v_0p=-\xi_{p+Q\up}$. The summation $\sum_p^\prime$ is
restricted in the region $-k_F<p<k_F$ as $-2k_F<k<0$ and
$0<k+Q<2k_F$. Introducing a Bogoliubov transformation,
\be
\eta_{p} = u_p c_{p\down} - v_p c_{p+Q\up}, ~~
\lambda_{p} = v_p c_{p\down} + u_p c_{p+Q\up},
\ee
with
\be
u_p = \frac{1}{\sqrt{2}}\sqrt{1+\frac{\xi_{p\down}}{\sqrt{\xi_{p\down}^2+B^2}}},~~
v_p = \frac{1}{\sqrt{2}}\sqrt{1-\frac{\xi_{p\down}}{\sqrt{\xi_{p\down}^2+B^2}}},
\ee
the Hamiltonian is diagonalized to be
\be
H_{\rm MF} = \sum_{-k_F<p<k_F}
E_p(\eta_p^\dagger\eta_p-\lambda_p^\dagger\lambda_p) + 2US_0^2,
\ee
with $E_p = \sqrt{\xi_{p\down}^2+B^2}$. A gap is opened at the Fermi
surface and the system becomes an insulator. Via the variational
method, one obtains the gap equation,
\be
\frac{1}{U} =
\sum_{-k_F<p<k_F}\left(\frac{1}{E_p}+\frac{\partial}{\partial E_p}\right)\left[n_F(-E_p)-n_F(E_p)\right],
\ee
which determines $S_0$.

\begin{figure}[t]
\includegraphics[height=3.75cm]{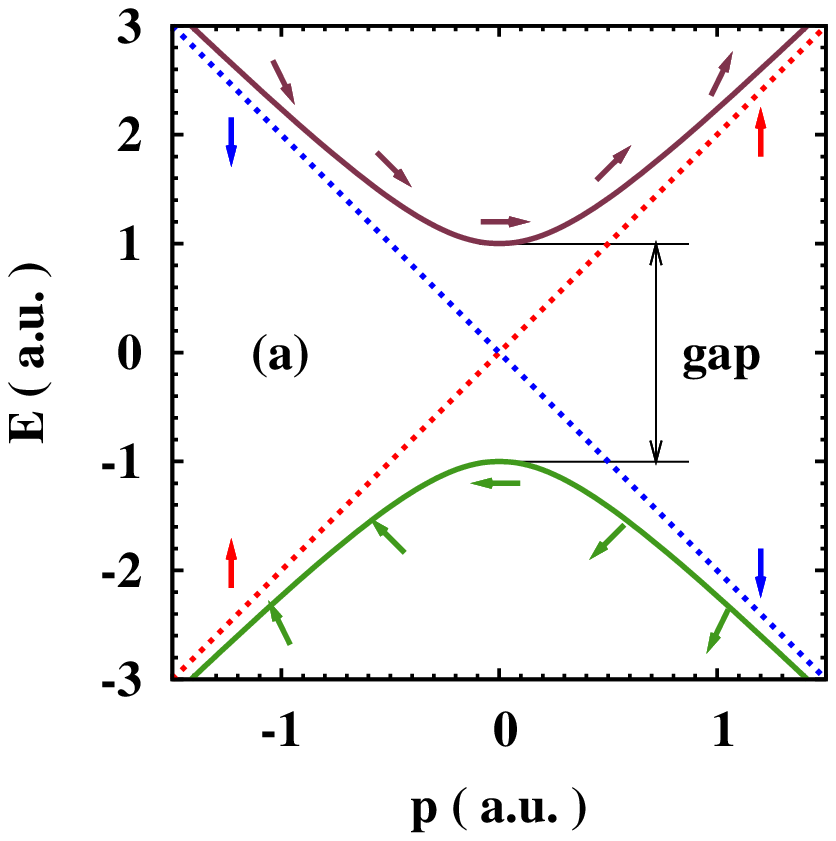}\includegraphics[height=3.75cm]{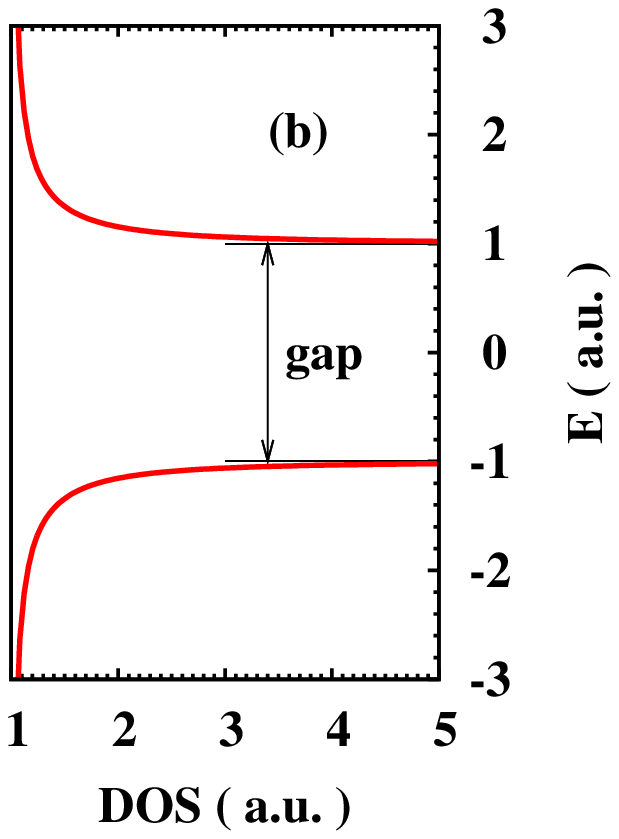}
\caption{ (Color online) (a) Spectrum $\pm E_p$ (full curves) and spin configuration
  of quasi-particle in the helical SDW state compared with the initial
  spectrum $\xi_{p\down}$ and $\xi_{p+Q\up}$ (dotted curves) and spin
  configuration. (b) Density of states of quasi-particle in the helical SDW state.}
\end{figure}

The quasi-particle in the helical SDW phase are described by
$\eta_p$ and $\lambda_p$ with $-k_F<p<k_F$. The spectrum and density
of states are plotted in Fig.~4. It is indicated that the gap
opens at the Fermi surface. The spin states of quasi-particles
are no longer pure spin-up or spin-down, but mixed spin state,
{\em enabling back scattering}.

\subsection{Surfaces of 3D TIs with Hexagonal Fermi Surface}

We first need to determine {\em which} spin density wave emerges in
the SDW state. Following the Ginzburg-Landau theory, the free-energy
functional can be written as 
\be
F_{GL}=F_2 + F_{4h}
\ee
with
\be
F_2=\frac{1}{2}\sum_{\nu,{\bf q}}\tilde{\chi}_{\nu,\nu}^{-1}({\bf
  q},0)|\phi_\nu({\bf q})|^2.
\ee
Here $\tilde{\chi}_{\nu,\nu}$ is the
diagonalized and ``normalized'' [see Eq.~(\ref{chi-G})] spin
susceptibility tensor. $\phi_\nu({\bf q})$ is the SDW. $F_{4h}$
denotes the free-energy to the fourth and higher orders
of $\phi$. At sufficient low temperature, $\chi$ and $F_2$ become
negative, which drives the system into the SDW state. As
\be
\tilde{\chi}_{\nu,\nu}^{-1}=\frac{1-U\tilde{\chi}_{\nu\nu}^0}{\tilde{\chi}_{\nu\nu}^{0}}=-U+1/\tilde{\chi}_{\nu\nu}^0,
\ee
the SDW with largest $\chi^{0}_{\nu\nu}$ will first emerge as
temperature is lowered down. According to the discussion in Sec.~III~B,
such spin susceptibility is $\tilde{\chi}_G^0({\bf Q}_i,0)$ [$i=(1,2,3)$].
Therefore, the first emergent SDW is the {\em helical} SDW. As
$\tilde{\chi}_G^0({\bf Q}_i,0)$ is much larger than the others, there will
be a temperature region, where only the helical SDW emerges.
The following discussion is restricted in this temperature region.\cite{note2}

Free energy $F_2$ of the concerned helical SDWs is,
\be
F_2=\frac{1}{2}\sum_i\tilde{\chi}_G^{-1}({\bf Q}_i,0)|\phi_i|^2,
\ee
with 
\be
\phi_i\equiv\ave{\tilde{\sigma}_G({\bf Q}_i)}
\ee being complex.
According to Eq.~(\ref{Gsym}), $F_{GL}$ should be invariant
under the symmetry operations: 
\be
\phi_i\to\phi_{i+1},\quad \phi_i\leftrightarrow
-\phi_i^\ast.
\ee The $C_{3v}$ symmetry also gives another invariant
operation,
\be
\phi_i\to \phi_{i-1}.
\ee
Besides, $F_{GL}$ respects the translational symmetry,
\be
{\bf r}\to {\bf r}+{\bf d}: \phi_i\to 
\phi_ie^{\ii{\bf Q}_i\cdot{\bf d}}.
\ee
Therefore, the symmetry allowed terms in $F_{GL}$ at the fourth order are, 
\be 
F_{4}=\sum_i u_{41}|\phi_i|^4+u_{42}|\phi_i|^2|\phi_{i+1}|^2.
\ee
To the sixth order,
\bea
F_{6}&=&\sum_i u_{61}|\phi_i|^6 +
u_{62}(|\phi_i|^4|\phi_{i+1}|^2 + |\phi_i|^2|\phi_{i+1}|^4)
\nonumber\\
&&\mbox{}~ + u_{63}[(\phi_1\phi_2\phi_3)^2+(\phi_1^\ast\phi_2^\ast\phi_3^\ast)^2],
\eea
where the last term appears due to the translational sysmmetry and
$\sum_i {\bf Q}_i=0$.\cite{Fu} Those higher order terms determines
whether the three SDWs coexist or only one of the SDWs is allowed
(the stripe phase). Explicitly, after reorganization, there are terms
like $(|\phi_{i}|^2-|\phi_{i+1}|^2)^{2n}$ with $n\ge 1$, which prefer
the difference $|(|\phi_{i}|^2-|\phi_{i+1}|^2)|$. If those terms are
prominent, the stripe phase is favored.

Below, we explore a mean field theory to describe the helical SDW
states on the surfaces of 3D TIs. For simplicity, we restrict ourself
to the stripe phase, where only one helical SDW exists. The
mean field Hubbard Hamiltonian is
\be
H_U = - \frac{U}{a_G} \sum_i  \left[ \tilde{\sigma}_G({\bf Q}_i)
  \phi_i^\ast + (\tilde{\sigma}_G({\bf Q}_i))^\dagger\phi_i - |\phi_i|^2\right] ,
\ee
The magnitude $|\phi_i|$ is determined by minimizing the
Ginzburg-Landau free energy, whereas phase fluctuation of $\phi_i$
correspond to the gapless the spin wave excitations
(Goldstone modes). In the stripe phase, only one of the three
$|\phi_i|$ is nonzero. The mean field Hamiltonian is
\bea
&&\hspace{-0.2cm} H_{\rm MF} = \frac{U}{a_G}\sum_i|\phi_i|^2 + \nonumber \\
&&\hspace{0.1cm} \sum_{{\bf k},i}{}^\prime\left(\hspace{-0.1cm}\begin{array}{cc}
  c^\dagger_{{\bf k}+} &
  c^\dagger_{{\bf k+Q}_i+} \end{array}\hspace{-0.1cm}\right) \left(\begin{array}{cc}
    \hspace{-0.1cm} \xi_{{\bf k}+} & \hspace{-0.1cm} B_i g_i({\bf k}) \\ \hspace{-0.1cm}
     B_i^\ast g_i^{\ast}({\bf k}) & \hspace{-0.1cm} \xi_{{\bf k+Q}_i+}
    \end{array}\hspace{-0.1cm}\right)\left(\hspace{-0.1cm} \begin{array}{c}
  c_{{\bf k}+} \\ c_{{\bf k+Q}_i+} \end{array}\hspace{-0.1cm}\right),\nonumber \\
\eea
with 
\be
g_i({\bf k})=\langle u_+({\bf k})|\tilde{\sigma}_G({\bf Q}_i)
|u_+({\bf k+Q}_i)\rangle,\quad B_i=-\frac{U}{a_G}\phi_i^\ast.
\ee
The summation $\sum_{\bf k}^\prime$ is restricted in certain region as
${\bf k}$ and ${\bf k+Q}_i$ dictate the same Hamiltonian. For example,
for the case where only $\phi_1$ is nonzero, ${\bf k}$ is restricted
in the region where $-Q_1\le k_x<0$ as ${\bf Q}_1=(Q_1,0)$. A
significant feature of the Hamiltonian is that both the diagonal and
off-diagonal terms are ${\bf k}$-dependent, which may induce
nontrivial Berry curvature in some cases. But we will not discuss this
important property here. The mean field Hamiltonian is diagonalized as 
\be
H_{\rm MF} = \sum_{{\bf k},i}{}^{\prime}
\left[E_{i{\bf k}+}\eta_{i{\bf k}}^\dagger\eta_{i{\bf k}}+E_{i{\bf
      k}-}\lambda_{i{\bf k}}^\dagger\lambda_{i{\bf
      k}}\right] + \frac{U}{a_G}\sum_i |\phi_i|^2 ,
\ee
via the following Bogoliubov transformation,
\bea
\eta_{i{\bf k}} &=& u_{i{\bf k}}e^{-\ii\psi_{i{\bf k}}} c_{{\bf k}+} - v_{i{\bf k}} c_{{\bf k+Q}_i+}\nonumber \\
\lambda_{i{\bf k}} &=& v_{i{\bf k}}e^{-\ii\psi_{i{\bf k}}} c_{{\bf k}+} + u_{i{\bf k}} c_{{\bf k+Q}_i+},
\eea
with
\bea
u_{i{\bf k}} &=& \frac{1}{\sqrt{2}}\sqrt{1+\frac{\zeta_{i{\bf k}}}
{\sqrt{\zeta_{i{\bf k}}^2+|B_ig_i({\bf k})|^2}}},\nonumber
\\
v_{i{\bf k}} &=& \frac{1}{\sqrt{2}}\sqrt{1-\frac{{\zeta_{i{\bf k}}}}
{\sqrt{\zeta_{i{\bf k}}^2+|B_ig_i({\bf k})|^2}}}.
\eea
and $\psi_{i{\bf k}}={\rm Arg}[B_ig_i({\bf k})]$, ${\zeta_{i{\bf
      k}}}=\frac{1}{2}(\xi_{{\bf k}+}-\xi_{{\bf k+Q}_i+})$. The 
energy spectrum of the quasi-particle excitation is
\be
E_{i{\bf k}\pm} = \pm \sqrt{\zeta_{i{\bf k}}^2+|B_ig_i({\bf k})|^2} 
+ \frac{1}{2}(\xi_{{\bf k}+}+\xi_{{\bf k+Q}_i+}).
\ee
It should be mentioned that although the energy gap is
${\bf k}$-dependent, it does {\em not} close at any ${\bf k}$.

\section{Helical magnetic order in the magnetically doped surfaces of 3D TIs}

The magnetic order in the magnetically doped surfaces of 3D TIs has
attracted a lot of
interest.\cite{MnDoped1,MnDoped2,Wray,QinLiu,aniso,Abanin} 
Theoretical investigation\cite{QinLiu} has shown that the
Ruderman-Kittel-Kasuya-Yosida (RKKY) interaction is always
ferromagnetic when the Fermi energy is close to the Dirac point.
When a ferromagnetic order emerges, a 
gap is opened around the Dirac point. Recent experiments confirmed
such results in Mn or Fe doped Bi$_2$Se$_3$ and
Bi$_2$Te$_3$.\cite{MnDoped1,MnDoped2,Wray}

Here we consider the situation when the Fermi energy is {\em much higher}
and the Fermi surface is hexagonal (and hence {\em nested}). From the
discussion in previous sections, we know that the spin susceptibility
function is peaked at the nesting wavevector where it is {\em
  helical}. Physically, the effective interaction between two magnetic
impurities are mediated by the carrier spin density excited by one of
the impurity and feeled by another. The spin susceptibility function
describes such excitation. The nature of the spin susceptibility thus
has profound implication on the effective interaction and the magnetic
order.

The above picture can be described well by the mean field Zener theory,
which has been shown to be successful in the theory of dilute III(Mn)-V
magnetic semiconductors.\cite{rmp} The Hamiltonian of the system is
\be
H = H_{STI} - J\sum_{iI}{\bf S}_I\cdot{\bf s}_i ,
\ee
where $H_{STI}$ is the Hamiltonian of carriers in the surfaces of
3D TIs. The last term describes the carrier--magnetic-impurity exchange
interaction. $I$ and $i$ label magnetic impurities and carriers
respectively (with spin ${\bf S}_I$ and ${\bf s}_i$ respectively).

In Zener theory, the equilibrium mean carrier and magnetic-impurity
spin densities are calculated by minimizing the Ginzburg-Landau free
energy. Under the mean field approximation and neglect
higher order correlations, the free energy functional is\cite{tomas} 
\bea
F_{GL}[\phi] &=& -n_Mk_BT \ln\left[\sum_{j=-S}^S\exp(J\phi j/Lk_BT)\right] +
F_c[\phi]\nonumber\\ && \mbox{}~- J\phi SB_S(SJ\phi/Lk_BT),
\label{EqGl}
\eea
where $\phi=\sqrt{\sum_\mu|\phi_\mu|^2}$ is the mean carrier spin
density. $L$ is the width of surface channel. The first term in right
hand side is the free energy of magnetic impurities with $S$ being
the total spin quantum number of a single impurity. $n_M$ denotes the
density of magnetic impurities. The second and third terms are the
free energy of the carrier system. The last term is the energy of
carrier spin density under the mean field of the exchange interaction,
with $B_S$ being the Brillouin function. 

The underlying physics is that, the increase in carrier spin density
$\phi$ reduces the first and last terms in the Ginzburg-Landau free
energy, whereas it costs by increasing $F_c[\phi]$ as the carrier
system is in the paramagnetic state. The equilibrium value of $\phi$
is determined by minimizing $F_{GL}[\phi]$. At small carrier spin
density (the ``linear susceptibility regime''), $F_c[\phi]$ can be
written as
\be
F_c[\phi] = \frac{1}{2L}\sum_{\nu,{\bf q}}\tilde{\chi}_{\nu,\nu}^{-1}({\bf q}) |\phi_\nu({\bf q})|^2
\ee
where $\tilde{\chi}_{\nu\nu}$ is the diagonalized and ``normalized''
[see Eq.~(\ref{chi-G})] spin susceptibility. It is noted that the
spin density $\phi_\nu({\bf q})$ corresponding to the largest
$\tilde{\chi}_{\nu\nu}({\bf q})$ is energetically favored as it minimizes
$F_c[\phi]$. In previous sections, we have shown that the largest 
spin susceptibility is achieved at $\tilde{\chi}_{G}({\bf Q}_i)$.
[see Fig.~3], where the corresponding $\phi_G({\bf Q}_i)$ is a 
{\em helical} spin density wave. Therefore, the energetically favored
magnetic order is the {\em helical magnetic order}, when Fermi surface
is nested.

It should be pointed out that the above equations can also be used to
obtain the Curie temperature of the magnetic order \cite{tomas,rmp} 
\be
k_B T_C = \frac{S(S+1)}{3}\frac{J^2 n_M \tilde{\chi}_G({\bf Q}_i,0)}{L}.
\ee
The mean carrier spin density is a stripe helical one at the nesting
wave-vector ${\bf Q}_i$. The spins of magnetic impurities are aligned
parallel or anti-parallel to the carrier spin density at their local
positions, depending on the sign of the exchange constant
$J$. Explicitly, the mean carrier spin density is
\be
\ave{\sigma_\mu({\bf r})} = \sum_i \ave{\sigma_{\mu}({\bf Q}_i)} e^{\ii {\bf Q}_i\cdot{\bf r}}
+ c.c. ,
\ee
where
\be
\ave{\sigma_{\mu}({\bf Q}_i)} =
\frac{{G}_{i\mu}}{\sqrt{a_G}} \phi_i ,
\ee
with ${\mu}=(x,y,z)$ and $\phi_i=\ave{\tilde{\sigma}_G({\bf Q}_i)}$ is
complex. Here $a_G$ is defined in Eq.~(\ref{a_G}) and ${G}_{i\mu}$ is
defined in Eq.~(\ref{Gi}) [the special case $i=1$ is given
approximately in Eq.~(\ref{Gmu})].

The amplitude $\phi=|\phi_i|$ is determined by minimizing
$F_{GL}[\phi]$. There are three energetically favored configurations
corresponding to the three ${\bf Q}_i$. For each
configuration, the magnetic anisotropy is expected to be {\em very
  large}, as the spin susceptibility tensor $\hat{\chi}({\bf Q}_i,0)$
is highly anisotropic [it has an eigen-value much larger than the
  other two, see Fig.~2(b)].

According to our calculation, for Mn doped Bi$_2$Te$_3$ surface states
with $E_F\simeq 0.7E_s$ (0.2~eV, corresponding electron density $4\times
10^{12}$~cm$^{-2}$), $\tilde{\chi}_G\simeq 0.3*10^{-2}$~\AA$^{-2}$eV$^{-1}$
[see Fig.~3(a)], $J\simeq 10^2$~eV~\AA$^{3}$,\cite{Dai} the width of
surface channel $L=50$~\AA~ (inferred from Ref.~\onlinecite{YLChen}),
$n_M=2\times 10^{-3}$~\AA$^{-3}$($\simeq 0.1$ mole fraction), $S=5/2$,
we get $T_C\simeq 30$~K, which is not very low.

It should be pointed out that further investigations on the problem
are demanded. On one hand, the carrier--magnetic-impurity exchange
interaction, which leads to carrier spin relaxation and shortens the
propagating distance of the carrier SDW excited by
magnetic impurities, should be included in the
calculation of the spin susceptibility function. On the other hand,
the exchange and correlation corrections to the spin susceptibility
function should be included.\cite{rmp} A density-functional
calculation will be appreciated.\cite{rmp} Via such improvement, the
magnetization can be calculated at arbitrary temperature, carrier and
magnetic-impurity densities. We believe that the helical magnetic
order is still favored after those corrections are included, according
to the symmetry of the system. We assumed that the electron density
can be tuned either by gate-voltage or doping by other dopants besides
ferromagnetic impurities, which is in principle achievable in
experiments.\cite{rmp}

Finally, we note that very recently, Ye et al. found that the helical
magnetic order emerges in a chain of impurity spins in the surfaces of
3D TIs.\cite{spin-helix} And Zhu et al. discussed the
RKKY interaction when Fermi surface is hexagonal.\cite{Zhu}

\section{Conclusion}

We study spin susceptibility and magnetic order at the edges/surfaces
of topological insulators when the Fermi surface is nested. We find
that due to spin-momentum locking as well as time-reversal symmetry,
spin susceptibility at the nesting wavevector has a strong {\em
  helical} feature. It follows then, a {\em helical} SDW
state emerges at low temperature due to Fermi surface
nesting. The helical feature of spin susceptibility also has
profound impact on the magnetic order in the magnetically doped
surfaces of 3D TIs. From the Zener theory, to the lowest order, we
predicted a {\em helical magnetic order} in such system.

The helical SDW order can be probed/determined either directly by spin
resolved STM or indirectly by the existence of an energy gap at the
Fermi surface via, e.g., ARPES. For spin pump-probe measurements, if a
local spin density is excited, it will propagator with certain
helicity along the nesting wavevectors. The helical magnetic order in
magnetically doped surface of 3D TIs can also be probed directly by
spin resolved STM and other magnetic response measurements.

Finally, recent studies indicate that topological insulators can
also be achieved in cold-atom systems.\cite{cold-atom} The tunability
of inter-particle interaction (and many other properties) in such
systems may be utilized to enhance the helical SDW order predicted
in this paper.

\section*{Acknowledgements}
Work at the Weizmann Institute  was supported by the German Federal
Ministry of Education and Research (BMBF) within the framework of the
German-Israeli project cooperation (DIP) and by the Israel Science
Foundation (ISF). We thank M. A. Martin-Delgado for
discussions. J.H.J. thanks M. Dolev for bringing
Ref.~\onlinecite{Alpich} to his attention. S.W. thanks Canadian NSERC
for support.


\begin{thebibliography}{99}
\bibitem{note0} If exchange and correlation effects are taken
          into account, Coulomb interaction can also affect spin
          susceptibility, which is beyond the scope of this paper.
\end{thebibliography}
\end{document}